\documentclass{PoS}
\usepackage{graphicx}
\usepackage{colordvi}
\usepackage{multirow}

\title{Spectrum of closed $k$-strings in D=2+1.}

\ShortTitle{Spectrum of closed $k$-strings in D=2+1}

\author{\speaker{Andreas Athenodorou} \\Rudolf Peierls Centre for Theoretical Physics \\
        University of Oxford\\
        E-mail: \email{athinoa@thphysics.ox.ac.uk}}

\author{Barak Bringoltz \\ Physics Department\\
        University of Washington, Seattle, WA 98195-1560, USA\\
        E-mail: \email{barak@phys.washington.edu}}

\author{Michael Teper  \\ Rudolf Peierls Centre for Theoretical Physics \\
        University of Oxford\\
    E-mail: \email{m.teper1@physics.ox.ac.uk}}

\abstract{We calculate the excitation spectrum of closed $k$-strings in 2+1 dimensional $SU(N)$ gauge theories for $N=4$, $5$ and $k=2$. Our results demonstrate that the low-lying spectrum of the $k=2$ string falls into sectors that belong to nearly pure antisymmetric and symmetric representations, showing that $k$-strings know not only about the centre of the group, but also about the full group. We also observe that the lightest states in each irreducible representation are consistent with what one would expect from an effective string theory that belongs to the same bosonic universality class (Nambu-Goto) as the fundamental string. We find that the corrections compared to the free string theory are of ${\cal O}(1)$, in striking contrast to the very small correction observed for the fundamental string. We also observe unbound $w=2$ states.}

\FullConference{The XXVI International Symposium on Lattice Field Theory\\
         July 14-19 2008\\
         Williamsburg, Virginia}

\begin{document}
\section{Introduction}
\vskip -0.5cm
The study of $k$-strings has been an active research topic for many years, involving several fields of research such as AdS/CFT, MQCD, Hamiltonian approaches and Lattice. The investigation of these flux tubes in several representations using lattice techniques is important for a number of reasons. First of all, we would like to know what effective string theory model describes the confining flux tube at any representation ${\cal R}$. Secondly, lattice calculations provide significant results that might help people studying $k$-strings in diverse aspects to aim for the appropriate model. Our attempt to study the $k=2$ spectrum has been motivated by the fact that the non triviality of the excitation spectrum will provide us with significant information about the effective string theory model that describes these states and moreover that the excitation spectrum of such observables has not yet been explored systematically due to the complexity of these calculations.

In our last paper \cite{ABM} we showed that the spectrum of closed tubes of fundamental flux ($k=1$) can be very well described by Nambu-Goto~(NG) effective string theory \cite{NGpapers} in flat space-time. What we actually showed is that our data is consistent with NG down to very short lengths, comparable to the physical length scale $\sim 1 / \sqrt{\sigma_f}$, where $\sigma_f$ is the string tension of the fundamental string. This is in striking contrast with what one observes if compares the data to the L\"uscher \cite{LW_old} and L\"uscher and Weisz \cite{LW_new} predictions. Apart from the energy level agreement, NG also predicts the degeneracy pattern of the observed closed fundamental strings, showing that they are really described by an effective string theory that belongs to the same universality class as NG. 

The higher representation closed flux-tubes under investigation have lengths $l$ that range from $l \sqrt{\sigma_f} \sim 1.5$ ${\rm to}$ $4.5$. We study the gauge groups $SU(4)$ and $SU(5)$ for $\beta=50.000$ and $80.000$ respectively, which corresponds to common value of the lattice spacing, $a \simeq 0.06 \ {\rm fm}$ using the convention $\sigma \equiv (440{\rm MeV})^2$. Our work consists of three different calculations. In the first part we project onto the $k=2$ antisymmetric and symmetric representations. In the second part we calculate the spectrum of $k=2$ strings without projection (i.e. using operators of type: ${\rm Tr}\{ U \}^2$ and ${\rm Tr}\{ U^2 \}$), and in the last part we calculate a potentially larger spectrum using the operators from the second part and a new set of operators which are designed to give rise to unbound doubly wound ($w=2$) states. 

During the last decade a lot of effort has been invested by the lattice community in studying confining flux tubes. This includes the investication of both open and closed $k-$strings in different representations. We refer the reader to some recent papers \cite{strings-lattice-review} and references therein. For more details on the calculation and relevant other references see our longer forthcoming write-up \cite{ABM2}.
\vspace{-0.65cm}
\section{Lattice Calculation: General}
\vskip -0.5cm
Our gauge theory is defined on a three-dimensional Euclidean space-time lattice that has been toroidally compactified with $L \times L_{\perp} \times L_T$ sites. The size of the string is equal to $L$; $L_{\perp}$ and $L_T$ are carefully chosen to be large enough in order to avoid any finite volume effects. For the spectrum calculation we perform Monte-Carlo simulations using the standard Wilson plaquette action:
\vskip -0.6cm
\begin{eqnarray}
S_{\rm W}=\beta \sum_P \left[ 1- \frac{1}{N} {\rm Re} {\rm Tr}{U_P} \right],
\label{eq:SW}
\end{eqnarray}
\vskip -0.4cm
The bare coupling $\beta$ is related to the dimensionful coupling $g^2$ through $\lim_{a\to 0}\beta={2N}/{ag^2}$. In the large--$N$ limit, the 't Hooft coupling $\lambda=g^2N$
is kept fixed, and so we must scale $\beta=2N^2/\lambda \propto N^2$ in order to keep the lattice spacing fixed.
The simulation we use combines standard heat-bath and over-relaxation steps in the
ratio 1 : 4. These are implemented by updating $SU(2)$ subgroups using the Cabibbo-Marinari algorithm.
\vspace{-0.7cm}
\section{Lattice Calculation: Operators}
\vskip -0.4cm
Since we are interested in the excitation spectrum of $k=2$ strings, it is necessary to find a way to project onto such states. The way to achieve this is to find a suitable basis described by good quantum numbers in which our operators will be encoded. In our case, this includes parity $P$, which shows how a string transforms under reflections over the longitudinal plane defined by the longitudinal and temporal lattice directions. This suggest that we need to introduce transverse deformations in Polyakov loops, construct line paths that transform in a certain way under such reflections and then use the variational technique to extract the spectrum. In general, the more operators we use the better the results we obtain; we therefore construct a plethora of Polyakov paths trying to extract states with high overlaps. 

We have also attempted to check whether the $k=2$ string spectrum includes unbound $w=2$ states, by using appropriate operators. These new states are expected to be described by frequencies lower than those describing the $k=2$ bound states. Some $k=2$ operators look like they wind twice around the torus, such as Eq.~(\ref{k2}). However, the way we had been constructing them prohibits us to project onto states with lower frequencies, since each Polyakov loop starts and ends at the same lattice point within one lattice size. To this purpose we construct Polyakov lines that wind twice around the torus with transverse deformations at the joint of the two lattices, as in Eq.~(\ref{w2}). We use the square-pulse deformation as an example to demonstrate how our operators are built:

\noindent
\underline{$1^{\rm st}$ set} of $k=2$ operators: (here and below the $\pm$ signs determine the parity of $\phi$)
\vspace{-0.3cm}
\begin{eqnarray}
\phi=\rm{Tr} \ \left\lbrace \ \parbox{1cm}{\rotatebox{0}{\includegraphics[width=1cm]{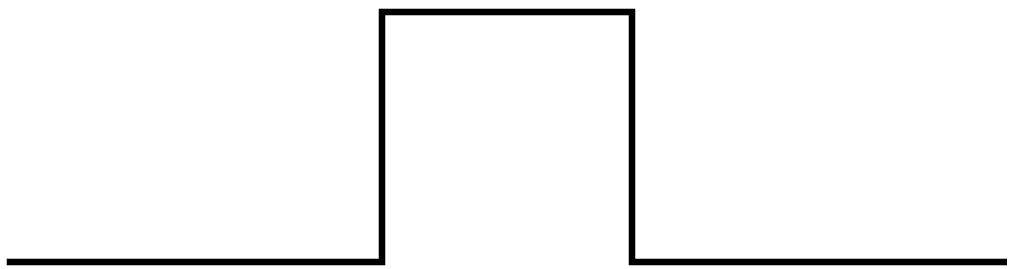}}} \ \right\rbrace^2 \pm \rm{Tr}  \left\lbrace \ \parbox{1cm}{\rotatebox{180}{\includegraphics[width=1cm]{square1w.eps}}} \ \right\rbrace^2
\label{k1}
\end{eqnarray}
\vspace{-1cm}

\noindent
\underline{$2^{\rm nd}$ set} of $k=2$ operators:
\vspace{-0.3cm}
\begin{eqnarray}
\phi=\rm{Tr} \ \left\lbrace \ \parbox{1cm}{\rotatebox{0}{\includegraphics[width=1cm]{square1w.eps}}} \cdot \parbox{1cm}{\rotatebox{0}{\includegraphics[width=1cm]{square1w.eps}}}  \ \right\rbrace \pm \rm{Tr}  \left\lbrace \ \parbox{1cm}{\rotatebox{180}{\includegraphics[width=1cm]{square1w.eps}}} \cdot  \parbox{1cm}{\rotatebox{180}{\includegraphics[width=1cm]{square1w.eps}}}  \ \right\rbrace
\label{k2}
\end{eqnarray}
\vspace{-0.8cm}

\noindent
\underline{$3^{\rm rd}$ set} of $k=2$ operators that are expected to project onto $w=2$ unbound states:
\vspace{-0.3cm}
\begin{eqnarray}
\phi=\rm{Tr} \ \left\lbrace \ \parbox{2cm}{\rotatebox{0}{\includegraphics[width=2cm]{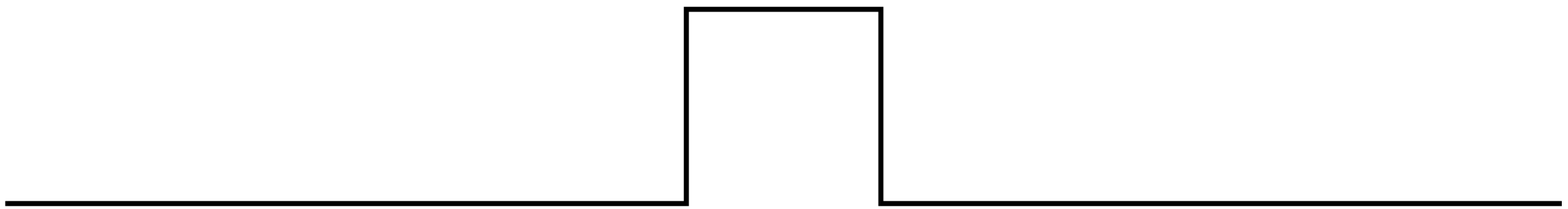}}} \ \right\rbrace \pm \rm{Tr}  \left\lbrace \ \parbox{2cm}{\rotatebox{180}{\includegraphics[width=2cm]{square2w.eps}}} \ \right\rbrace
\label{w2}
\end{eqnarray}
\vspace{-0.8cm}

\noindent
Projecting onto the \underline{$k=2$ antisymmetric representation}: 
\vspace{-0.3cm}
\begin{eqnarray}
\phi= [ \rm{Tr} \ \left\lbrace \ \parbox{1cm}{\rotatebox{0}{\includegraphics[width=1cm]{square1w.eps}}} \ \right\rbrace^2 - \rm{Tr} \ \left\lbrace \ \parbox{1cm}{\rotatebox{0}{\includegraphics[width=1cm]{square1w.eps}}} \cdot \parbox{1cm}{\rotatebox{0}{\includegraphics[width=1cm]{square1w.eps}}}  \ \right\rbrace ] \pm [ \rm{Tr}  \left\lbrace \ \parbox{1cm}{\rotatebox{180}{\includegraphics[width=1cm]{square1w.eps}}} \ \right\rbrace^2 - \rm{Tr}  \left\lbrace \ \parbox{1cm}{\rotatebox{180}{\includegraphics[width=1cm]{square1w.eps}}} \cdot  \parbox{1cm}{\rotatebox{180}{\includegraphics[width=1cm]{square1w.eps}}}  \ \right\rbrace]
\label{symm}
\end{eqnarray}
\vspace{-0.8cm}

\noindent
Projecting onto the \underline{$k=2$ symmetric representation}: 
\vspace{-0.3cm}
\begin{eqnarray}
\phi= [ \rm{Tr} \ \left\lbrace \ \parbox{1cm}{\rotatebox{0}{\includegraphics[width=1cm]{square1w.eps}}} \ \right\rbrace^2 + \rm{Tr} \ \left\lbrace \ \parbox{1cm}{\rotatebox{0}{\includegraphics[width=1cm]{square1w.eps}}} \cdot \parbox{1cm}{\rotatebox{0}{\includegraphics[width=1cm]{square1w.eps}}}  \ \right\rbrace ] \pm [ \rm{Tr}  \left\lbrace \ \parbox{1cm}{\rotatebox{180}{\includegraphics[width=1cm]{square1w.eps}}} \ \right\rbrace^2 + \rm{Tr}  \left\lbrace \ \parbox{1cm}{\rotatebox{180}{\includegraphics[width=1cm]{square1w.eps}}} \cdot  \parbox{1cm}{\rotatebox{180}{\includegraphics[width=1cm]{square1w.eps}}}  \ \right\rbrace]
\label{antsymm}
\end{eqnarray}
\vspace{-0.8cm}

\noindent
All the line paths used in the calculation can be found in \cite{ABM2}.
Our operators are also described by the quantum numbers of winding momentum $q = 0,\pm 1, \pm 2,\dots$ in units
of $2\pi/l$ and the transverse momentum which is set to zero. We also use the standard smearing/blocking technique for the Polyakov lines in order to enhance the projection of our operators onto the physical scales.

For each combination of the quantum numbers of $P$ and $q$ we construct the full
correlation matrix of operators and use it to obtain best estimates for the
string states performing a variational method applied to the transfer matrix $\hat{T}=e^{-aH}$.
\vspace{-0.5cm}
\section{Nambu-Goto String model}
\vskip -0.4cm
The action of Nambu-Goto is proportional to the surface area of the world sheet swept by the propagation of the string; for closed strings this world sheet is a torus. 
The NG states correspond to winding states characterised by the winding number $w$, which counts how many times the string wraps around the torus. The quantization of a Lorentz-invariant bosonic string is successful only in 26 dimensions due to the Weyl anomaly. It has been shown that this anomaly is suppressed for large enough strings \cite{Olesen} and so we can still think of NG model as an effective string theory for long strings in lower dimensions ($D<26$).

The NG energy spectrum corresponds to phonons travelling clockwise and anticlockwise along the closed string. The single-string states can be described by the winding number $w$, by the total contribution in energy from left and right moving phonons $N_L$ and $N_R$, by the centre of mass momentum which is set to zero $\vec p_{\rm c.m.}=0$, and finally by the momentum along the string axis in units of $2\pi q/l$ with $q=0,\pm 1,\pm 2,\dots$. The above quanta are not independent, and they obey the level matching constraint $N_L-N_R=qw$.

The string states can be characterised as irreducible representations of the $SO(D-2)$ group, which rotates the spatial directions transverse to the string axis. In our $D=2+1$ case this group becomes the transverse parity with eigenvalues $P=(-1)^{\rm number \ of \ phonons}$.
Finally, the energy of a closed-string state, described by the above quantum numbers for any $D$, is given by the following relation:
\vspace{-1.2cm}

\begin{eqnarray}
E^2_{N_L,N_R,q,w} &=& \left(\sigma \,l w\right)^2 + 8 \pi \sigma
\left( \frac{N_L+N_R}2 - \frac{D-2}{24}\right) + \left(\frac{2\pi
q}{l}\right)^2.\label{NG0}
\end{eqnarray}
\vspace{-0.7cm}

\noindent
Apart from the NG effective-string theory model obtained in a string theoretical context, other string theoretical effective models have been proposed. The first approach came up in the early eighties by L\"uscher for the case of $w=1$ and $q=0$ in \cite{LW_old}.  His result has been extended in \cite{LW_new}, where the authors used an open-closed string duality and imposed interaction terms to show that the spectrum of a closed flux tube in 2+1 dimensions is given by: 
\vspace{-0.5cm}

\begin{equation}
E^2_n = {\left( \sigma l \right)^2 + 8 \pi \sigma \left( n - \frac{1}{24} \right)} + {\cal O} \left(1/l^3 \right). \nonumber
\label{luscher2}
\end{equation}
\vspace{-0.5cm}

\noindent
Noting that the two first terms on the right hand side of Eq.~(\ref{luscher2}) are the predictions of the NG model for $q=0$ and $w=1$ (expanded to first order in $l\surd \sigma$), we use the following ansatz to fit our results and give empirical non universal corrections $C_p$ to the energy. 
\vspace{-0.4cm}
\begin{equation}
E^2_{\rm fit} = E^2_{NG} - \sigma
\frac{C_p}{\left(l\sqrt{\sigma}\right)^{p}}, \qquad p\ge
3\label{fit}
\end{equation}

\vspace{-0.8cm}
\section{Results: Symmetric and Antisymmetric representations}
\label{Results1}
\vskip -0.4cm
In this section we present our results from the projection onto the totally antisymmetric and symmetric representations.
The results have been obtained for $SU(4)$ with $\beta=50.000$ and for $SU(5)$ with $\beta=80.000$. 
The basis we have used consists of $\sim 80$ operators of the type: Eqs.~(\ref{symm}, \ref{antsymm}) and the energies were obtained using single cosh fits.

\begin{figure}[htb]
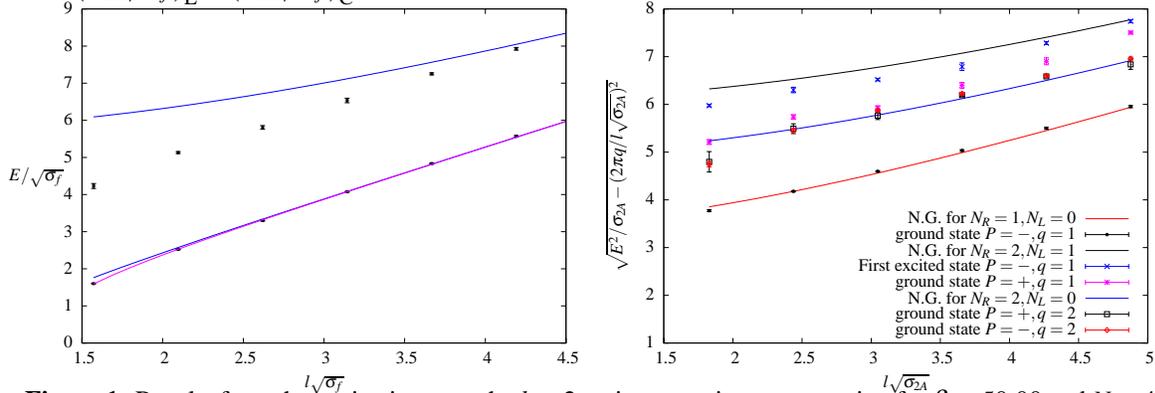

\centerline{ \scalebox{0.6}{\input{plotas.tex} \ \input{plot.tex}}}
\vskip -0.5cm
\caption{Results from the projection onto the $k=2$ antisymmetric representation for $\beta=50.00$ and $N=4$. Both plot parameters have been expressed in dimensionless units. \underline{Left panel:} The first two energy levels for $P=+$ and $q=0$ as a function of the string length.  The blue lines present the NG predictions for the ground and first excited states and magenta line presents the best fit onto the ground state which has been extracted using the fitting ansatz. \underline{Right panel:} The first five energy levels for $q=1,2$ as a function of the string length.} 
\label{figure1}
\vskip -0.4cm
\end{figure}

By focussing on the projection onto the antisymmetric representation, we realise that the ground state for $P=+$ and $q=0$ can be adequately described by NG with non universal corrections $C_3$ of order ${\cal O}(1)$. This is in striking contrast with what one observes for the fundamental representation, for which this correction is much smaller (see Table~\ref{table} and right panels of Fig.~\ref{Towers}). The first excited state for the same combination of quantum numbers seems to be approaching the NG prediction slowly with quite large deviations (left panels of Fig.~\ref{figure1}). 

It would be very interesting to see what can be obtained from the non-zero longitudinal momentum $q=1,2$ data. The lightest state has only one phonon $a_{-1} | 0>$ and is consequently described by $P=-$ and $q=1$. The right panel of Fig.~\ref{figure1} shows that this state is remarkably well described by the NG prediction (red line). The next two states (red and black) with $q=2$ also have a non-trivial phonon structure, $a_{-2} | 0>$ for $P=-$ and $a_{-1} a_{-1}| 0>$  for $P=+$, and are degenerate in the way NG predicts. The last two states with $q=1$ which have a much more complicated phonon structure with movers of both kinds attached on the vacuum, deviate a lot more from theoretical predictions. 

The general conclusion is that torelon states that correspond to string states with phonons of only one kind, either left or right moving, are well-approximated by NG. In contrast, torelon states with both kinds of movers have much higher deviations. For more details on the phonon structure of the string states, see our longer write up. Similar results with less accuracy have been observed for the symmetric representation. It is worth mentioning that states corresponding to the symmetric representation are massive, making the energy extraction an extremely difficult task.

We have also found that the string tension of the ground state for the symmetric~($2S$) and antisymmetric~($2A$) representations of $k=2$ string is remarkably close to the prediction of Casimir scaling. For example, for $SU(5)$ and $\beta=80.000$ we find that $\left( {\sigma_{2S}} / {\sigma_f} \right)_{\rm L}-\left( {\sigma_{2S}} /  {\sigma_f} \right)_{\rm C} \sim -3.2 \%$ and $\left( {\sigma_{2A}} / {\sigma_f} \right)_{\rm L}-\left( {\sigma_{2A}} / {\sigma_f} \right)_{\rm C} \sim +2.2 \%$ where L(C) stands for Lattice(Casimir).
\vskip -0.5cm
\vspace{-0.4cm}
\section{Results: $k=2$ Spectrum: Group $SU(N)$ or Center $Z_N$?}
\label{Results2}
\vskip -0.4cm
Previous research \cite{BM1} has shown that non-trivial bound $k$-strings really exist for $N<\infty$. The ground state of a $k$-string is stable and corresponds to the totally antisymmetric representation; this is the representation of ${\cal N}$-ality $k$ with the smallest string tension, in both the Casimir scaling and the MQCD scenarios. Flux tubes in different representations ${\cal R}$ will be unstable and screened down by gluons to the totally antisymmetric one. However, unstable strings are also expected to be visible, since they appear as nearly-stable excited states in the energy spectrum of the string if the amplitude of the gluon screening is small enough.  

To investigate whether the energy spectrum includes unstable flux tubes of representations different than the antisymmertic, we need to use a basis of operators of type: Eqs.~(\ref{k1}, \ref{k2}). Once more we perform calculations for $SU(4)$ with $\beta=50.000$ and for $SU(5)$ with $\beta=80.000$. By comparing our results to what we have obtained in Section ~\ref{Results1} we find that our $k=2$ spectrum can be described by two sectors, which belong to two different irreducible representations, the totally antisymmetric and the totally symmetric. We also observe that the ground state for each different configuration of the quantum numbers $P$ and $q$ is always antisymmetric, but the rest of the spectrum has a much more complicated structure, since the energy levels of the two different irreducible representations cross. 

\begin{figure}
\centerline{\hspace{20mm} \scalebox{0.65}{\input{plotnew3.tex} \ 
\put(-315,250){\large $l \sqrt{\sigma_f}\sim2.1$}
\put(-195,250){\large $l \sqrt{\sigma_f}\sim3.1$}
\put(-75,250){\large $l \sqrt{\sigma_f}\sim4.2$}
\put(-73,110){$\parbox{0.3cm}{\rotatebox{0}{\includegraphics[width=0.3cm]{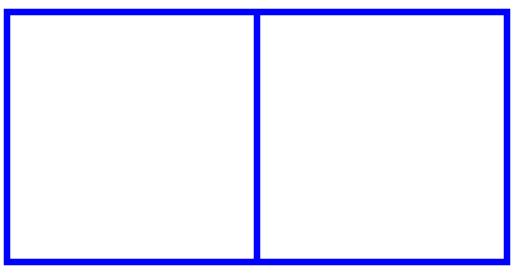}}}$}
\put(-24,110){$\parbox{0.3cm}{\rotatebox{270}{\includegraphics[width=0.3cm]{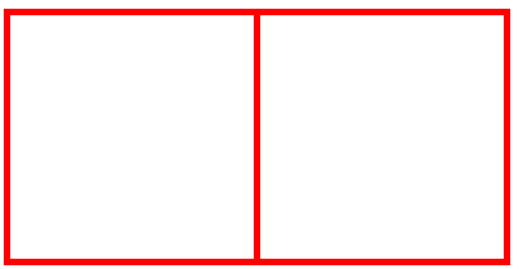}}}$}
\put(-194,85){$\parbox{0.3cm}{\rotatebox{0}{\includegraphics[width=0.3cm]{symmetrictab.eps}}}$}
\put(-144,85){$\parbox{0.3cm}{\rotatebox{270}{\includegraphics[width=0.3cm]{antisymmetrictab.eps}}}$}
\put(-314,60){$\parbox{0.3cm}{\rotatebox{0}{\includegraphics[width=0.3cm]{symmetrictab.eps}}}$}
\put(-264,60){$\parbox{0.3cm}{\rotatebox{270}{\includegraphics[width=0.3cm]{antisymmetrictab.eps}}}$}
\hspace{-30mm} 
\scalebox{1.0}{\input{plot1new.tex}}  \hspace{-70mm} 
\scalebox{1.0}{\input{plot2new.tex}} 
}}
\vskip -0.3cm
\caption{\underline{Three left panels:} Energy towers for three different string legths ($l\sqrt{\sigma_{f}} \sim 2.1, \  3.1, \ 4.2$fm) where we present the energies of the six lowest states in the $k=2$ spectrum of $SU(4)$ for $P=+$, $q=0$ and $\beta=50.000$ vs. the Overlap onto the particular irreducible representation. Data in Red(blue) corresponds to the projection onto the antisymmetric(symmetric) representation. \underline{Two right panels:} Ground states of the fundamental representation (second from the right) and $k=2$ antisymetric representation for $SU(5)$, $P=+$, $q=0$ and $\beta=80.000$. In red are the NG predictions and in blue are our fits.}
\label{Towers}
\vskip -0.3cm
\end{figure}

In the three left panels of Fig.~\ref{Towers}, we present the energy towers for three different string lengths. The towers show that each state is nearly entirely antisymmetric or symmetric with an overlap close to 1. The guide lines demonstrate how the energy levels alternate from being antisymmetric to being symmetric. Our results show clearly that the $k=2$ closed strings know about the full group theoretical structure of $SU(N)$ rather than just about its centre $Z_N$.
\vspace{-0.7cm}
\section{Results: Unbound $w=2$ states}
\vskip -0.5cm
In this section we present the results from our attempt to project onto unbound $w=2$ string states. We mentioned above that such states require new operators constructed in such a way that they will give rise to them. The way to check whether these states really exist is to perform the same calculation as in the previous section with an extended basis that includes the new operators, and then to compare the new results to the ones obtained 
before. This new extended basis consists of approximately $\sim 240$ operators.   

The calculation shows that by using this new extended basis of operators we get states that could not be seen before. We also observe that the use of these new operators enhances the overlap onto the other bound states and thus decreases their errors. 

\begin{figure}[htb]
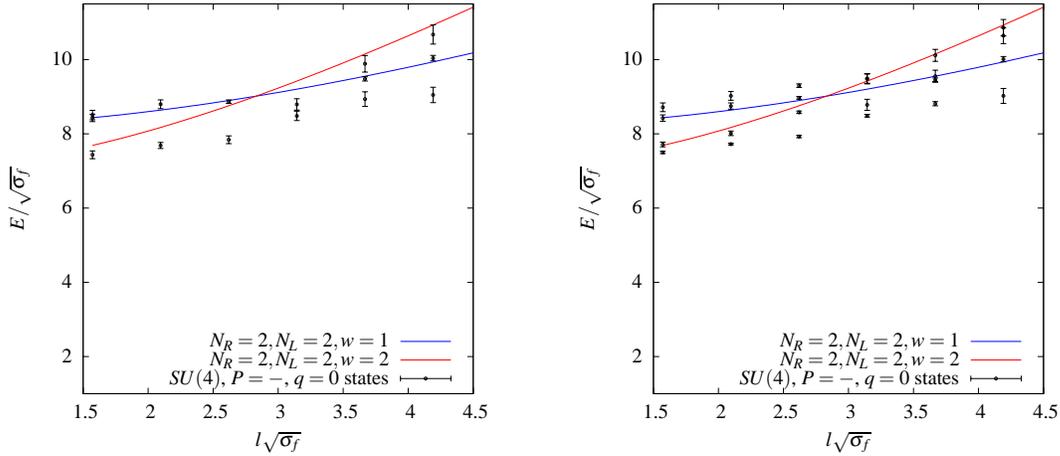

\centerline{\scalebox{0.70}{\input{plotnd.tex}} \hspace{-15mm}  \scalebox{0.70}{\input{plotd.tex} }}
\vskip -0.3cm
\caption{First few states for $SU(4)$, $P=-$, $q=0$, and $\beta=50.000$. In blue(red) we present the $w=1(2)$ NG prediction with $\sigma=\sigma_{2A}(\sigma_F)$. 
\underline{Left panel}: results with operators of type: Eqs.~(3.1, 3.2). \underline{Right panel}: results with operators of type: Eqs.~(3.1, 3.2, 3.3) -- it is clear that new states appear in the spectrum.}
\label{Extra}
\vskip -0.5cm
\end{figure}

In Fig.~\ref{Extra} we demonstrate that the new data for $SU(4)$, $P=-$ and $q=0$ indeed contain extra states. The red line presents the NG prediction for a double winding number string $w=2$ with the fundamental string tension $\sigma_f$. As can be seen, it is close to some of the new states appearing on the spectrum. This suggests that it is highly plausible that these new states belong to an effective string theory model in the same universality class as the double winding number NG string. Similar results have been observed for the case of $SU(5)$. 

\begin{table}
\centering{
\begin{tabular}{|c|c|c|c|} \hline \hline
 $ $  & $a^2 \sigma$ & $C_3$ & $\chi^2 / d.o.f$ \\  \hline \hline
$k=1$ for $SU(4)$ and $\beta=50.000$ & $0.017158(42)$ & $0.193(84)$ & $0.862$ \\ \hline
$k=2$ for $SU(4)$ and $\beta=50.000$ & $0.023218(39)$ & $2.64(18)$ & $1.535$ \\  \hline
$k=1$ for $SU(5)$ and $\beta=80.000$ & $0.016849(19)$ & $0.018(45)$ & $0.2415$ \\  \hline
$k=2$ for $SU(5)$ and $\beta=80.000$ & $0.025849(39)$ & $2.14(17)$ & $0.613$ \\  \hline
\end{tabular}}
\caption{Fits using equations 4.1 and 4.3.}
\label{table}
\vskip -0.4cm
\end{table}

\vspace{-0.2cm}
\section{Summary}
\vskip -0.2cm
We have calculated the energy spectrum of closed $k=2$ strings in 2+1 dimensions for $N=4$ and $5$ and $\beta=50.000$ and $80.000$ respectively. The calculation is divided in three parts. In the first part we project onto the $k=2$ symmetric and antisymmetric representations, in the second part we calculate the $k=2$ spectrum using the basis that includes operators of the type ${\rm Tr}\{ U \}^2$ and ${\rm Tr}\{ U^2 \}$ and in the third part we calculate the $k=2$ spectrum using the operators of the second part and in addition the operators that have been constructed in such a way to project onto unbound $w=2$ states. Each calculation has been performed for different parities and longitudinal momenta.

What we observe is that our $k=2$ spectrum falls into sectors that belong to two irreducible representations of $SU(N)$, namely the symmetric and the antisymmetric. This shows that $k$-strings know about the full gauge group and not just about its centre. We also demonstrate that the $k=2$ antisymmetric representation is clearly well described by the NG model and there is also evidence that so is the symmetric with larger deviations. Finally, we find that the $w=2$ spectrum has unbound states that can be accommodated in a model that belongs to the same universality class as the $w=2$ NG model.   

\vspace{-0.4cm}
\section*{Acknowledgements}
\vskip -0.45cm
The computations were carried out on EPSRC and Oxford funded computers in Oxford Theoretical Physics. AA acknowledges the support of the EC 6$^{th}$ Framework Programme Research and Training Network MRTN-CT-2004-503369 and the LATTICE 2008 organizers for supporting him financially. BB was supported by the U.S. DOE under Grant No. DE-FG02-96ER40956.

\end{document}